\documentclass[12pt]{article}
\usepackage{epsfig}
\newcommand{\be}{\begin{equation}}
\newcommand{\ee}{\end{equation}}
\newcommand{\bea}{\begin{eqnarray}}
\newcommand{\eea}{\end{eqnarray}}
\newcommand{\norsl}{\normalsize\sl}
\newcommand{\norsc}{\normalsize\sc}

\newcommand{\nn}{\nonumber}
\begin{document}

\begin{titlepage}

\title{Transition from real to virtual
\\ polarized photon structures}
\author{
\norsc  Takahiro UEDA$^{a)}$\thanks{e-mail address: t-ueda@phys.ynu.ac.jp}~, 
Tsuneo UEMATSU$^{b)}$\thanks{e-mail address: uematsu@phys.h.kyoto-u.ac.jp}~, 
Ken SASAKI$^{a)}$\thanks{e-mail address: sasaki@phys.ynu.ac.jp} \\
\norsl a) Dept. of Physics,  Faculty of Engineering, Yokohama National
University \\
\norsl  Yokohama 240-8501, JAPAN \\
\norsl b)~ Dept. of Physics,  Graduate School of Science,  Kyoto University \\
\norsl     Yoshida, Kyoto 606-8501, JAPAN \\
}

\date{}
\vspace{2cm}

\maketitle

\vspace{2cm}

\begin{abstract}
{\normalsize
We investigate the transition of the polarized photon structure 
function $g_1^\gamma(x,Q^2,P^2)$ when the target photon shifts 
from  on-shell ($P^2=0$) to far off-shell ($P^2\gg \Lambda^2$) region.  
The analysis is performed to the next-to-leading order in QCD.
The  first moment of $g_1^\gamma$ which vanishes for the real photon,  
turns to be a negative value when target photon becomes off-shell.
The explicit $P^2$-dependence of the first moment sum rule as well as 
of the structure function $g_1^\gamma(x,Q^2,P^2)$ as a function of $x$
is studied in the framework of the vector meson dominance model. 
}
\end{abstract}

\begin{picture}(5,2)(-290,-550)
\put(2.3,-75){YNU-HEPTh-06-104}
\put(2.3,-90){KUNS-1993}
\put(2.3,-105){June 2006}

\end{picture}

\thispagestyle{empty}
\end{titlepage}
\setcounter{page}{1}
\baselineskip 18pt
%
The photon structure has been investigated both theoretically 
and experimentally for decades \cite{KZS,Nisi,Klas,Schi}. 
And in recent years there has been growing interest in the
spin structure of the photon. In particular,
the first moment of the polarized photon structure function
$g_1^\gamma$ has attracted a lot of attention in the literature
\cite{ET}-\cite{BBS} due to its connection with the axial anomaly.
This structure function $g_1^\gamma(x,Q^2,P^2)$, where $q^2=-Q^2$  
($p^2=-P^2$) is the mass  squared of the probe (target) photon,    
can be measured in the experiments either of the polarized $ep$ collision
\cite{Barber,SVZ} or of  the polarized $e^+e^-$ collision in the future linear
collider  ILC~\cite{GSV,Shore}.

A unique and interesting feature  of the photon structure functions is
that, in contrast with the nucleon case, the target mass squared $-P^2$ 
is not fixed but can take various values and that the structure functions show 
different behaviors depending on the values of $P^2$. 
The QCD analysis of $g_1^\gamma$ for  a real photon ($P^2=0$) target was 
performed in the leading order (LO) \cite{KS}, and in the 
next-to-leading order (NLO) \cite{SVZ, GRS}. In the case of a
virtual photon target ($P^2\ne0$), $g_1^\gamma(x,Q^2,P^2)$ was
investigated  up to the NLO in QCD by the present authors in
\cite{SU1}, and  in the second paper of \cite{GRS}. 
Moreover, the polarized parton distributions inside the 
virtual photon were analyzed in various factorization schemes
\cite{SU2}, and the target mass effects of $g_1^\gamma(x,Q^2,P^2)$ was 
studied in \cite{SU3}. In Refs.\cite{SU1,SU2} the structure 
function $g_1^\gamma(x,Q^2,P^2)$ was analyzed  in the 
kinematic region
\be
\Lambda^2 \ll P^2 \ll Q^2~, \label{KineRegion}
\ee
where $\Lambda$ is the QCD scale parameter. The advantage in studying the 
virtual photon target in the kinematic region (\ref{KineRegion}) is that 
we can calculate structure functions by the perturbative method without any 
experimental data input \cite{UW}, which is contrasted with the case of a real photon
target where in the NLO there exist nonperturbative pieces \cite{BB}-\cite{DG}.

In the present paper we analyze the  transition of the polarized 
photon structure function $g_1^\gamma$ when the target photon shifts from 
on-shell ($P^2=0$) to far off-shell in the region (\ref{KineRegion}).
In fact, $g_1^\gamma$ exhibits an interesting $P^2$-dependence. At $P^2=0$, 
the structure function
$g_1^\gamma$ satisfies a remarkable sum rule~\cite{ET}-\cite{BBS}:
\be
\int^1_0 g_1^\gamma(x,Q^2) dx=0~. \label{1momentReal}
\ee
But when the target photon becomes off-shell, $P^2\ne0$, the first moment 
of the corresponding photon structure function $g_1^\gamma(x,Q^2,P^2)$ does 
not vanish any more. For the case $\Lambda^2 \ll P^2 \ll Q^2$, 
the first moment has been calculated up to the 
NLO~\cite{NSV,SU1}, and quite recently up to the 
next-to-next-to-leading order (NNLO) in QCD~\cite{SUU}. The NLO 
result is
\bea
&&\int_0^1 dx g_1^\gamma(x,Q^2,P^2)\nonumber\\
&&\quad =-\frac{3\alpha}{\pi}n_f\langle e^4\rangle\left[
\left(1-\frac{\alpha_s(Q^2)}{\pi}\right)
-\frac{n_f\langle e^2\rangle^2}{\langle e^4\rangle}
\frac{2}{\beta_0}
\left(\frac{\alpha_s(P^2)}{\pi}-\frac{\alpha_s(Q^2)}{\pi}\right)\right]~,
\label{1momentVirtual}
\eea 
with $\beta_0=11-2n_f/3$ being the one-loop QCD $\beta$ function. Here $\alpha$ 
($\alpha_s(Q^2)$) is the QED (QCD running) coupling constant, $n_f\langle e^4\rangle
=\sum_{i=1}^{n_f}e_i^4$ and $n_f\langle e^2\rangle=\sum_{i=1}^{n_f}e_i^2$ with 
$e_i$ being the electric charge of the active quark (i.e., the massless quark) with 
flavor $i$ in the unit of proton charge and $n_f$ the 
number of active quarks. In order to investigate the transition of $g_1^\gamma$
from  on-shell to  far off-shell region, 
we derive a formula which accommodates a unified description of both regions. 
And we explain the transition from the vanishing first moment sum rule (\ref{1momentReal})
at  $P^2 = 0$ to the non-vanishing sum rule (\ref{1momentVirtual}) for
off-shell $P^2\neq 0$. For the explicit $P^2$-dependence we resort to the 
vector-meson-dominance (VMD) model~\cite{JJS}. 

In the framework of the operator product expansion  supplemented by the
renormalization group method, 
the $n$-th moment of 
$g_1^\gamma(x,Q^2,P^2)$ is given as follows \cite{SU1}:
\be
\int_0^1dx x^{n-1}g_1^\gamma(x,Q^2,P^2)
=\sum_{j=\psi,G,NS,\gamma}C_n^j(Q^2/\mu^2,\bar{g}(\mu^2),\alpha)
\langle \gamma(p)|R_n^j(\mu^2)|\gamma(p)\rangle~,\label{BasicFormula}
\ee
where $|\gamma(p)\rangle$ is the \lq\lq target\rq\rq\
 virtual photon state with 
momentum $p$, 
$R_n^j$ and $C_n^j$ are the twist-2 spin-$n$ operators and their coefficient
functions with $\mu$ being the renormalization point. The indices $\psi,G,NS$ and $\gamma$ 
stand for singlet quark,
gluon, nonsinglet quark and photon, respectively.
The photon structure functions are defined in the lowest order of the 
QED coupling constant $\alpha$, and in this order 
we have $\langle\gamma(p)|R_n^\gamma(\mu^2)|\gamma(p)\rangle=1$~
for the photon matrix element of the photon operator $R_n^\gamma$. 

In the previous work~\cite{SU1,SU2} we took the renormalization point 
at $\mu^2=P^2$, where $P^2$ is much larger than $\Lambda^2$, 
so that we could calculate perturbatively the photon matrix elements of the hadronic 
operators $\vec{R}_n=(R_n^\psi, R_n^G, R_n^{NS})$~\cite{UW}.
In the present case, $P^2$ varies from the deeply virtual region 
down to $P^2=0$. Therefore we 
take the renormalization point at $\mu^2=Q_0^2$, where
$Q_0^2$ satisfies the condition
\be
\Lambda^2 \ll Q_0^2 \ll Q^2~. \label{KineRegionQ0}
\ee
Then
the $n$-th moment of $g_1^\gamma(x,Q^2,P^2)$ is expressed as (see Eq.(3.13) 
of Ref.\cite{SU1}), 
\begin{eqnarray}
\int_0^1 dx x^{n-1}g_1^\gamma(x,Q^2,P^2)
&=&\frac{\alpha}{4\pi}\vec{A}_n(Q_0^2;P^2)
\cdot{M}_n(Q^2/Q_0^2,\bar{g}(Q_0^2))\vec{C}_n(1,\bar{g}(Q^2))
\nonumber\\
&&+\vec{X}_n(Q^2/Q_0^2,\bar{g}(Q_0^2),\alpha)
\cdot\vec{C}_n(1,\bar{g}(Q^2))
+C_n^\gamma~,
\label{mom-ope}
\end{eqnarray}
where $\vec{A}_n=(A_n^\psi, A_n^G,A_n^{NS})$ denote the photon matrix elements
of the hadronic operators $\vec{R}_n$
renormalized at $\mu^2=Q_0^2$ and are defined as
\bea
\langle\gamma(p)|\vec{R}_n(\mu)|\gamma(p)\rangle|_{\mu^2=Q_0^2}=
\frac{\alpha}{4\pi}\vec{A}_n(Q_0^2;P^2)~.
\eea
The evolution factors ${M}_n$ and $\vec{X}_n$ are given in terms of
$T$-ordered exponential as~\cite{BB}
\bea
&&{M}_n(Q^2/Q_0^2,\bar{g}(Q^2))={T}\exp\Bigl[
\int_{\bar{g}(Q^2)}^{\bar{g}(Q_0^2)}dg \frac{\hat{\gamma}_n(g)}{\beta(g)}
\Bigr]~,\label{FactorMn}\\
&&\vec{X}_n(Q^2/Q_0^2,\bar{g}(Q^2),\alpha)
=\int_{\bar{g}(Q^2)}^{\bar{g}(Q_0^2)}dg\frac{\vec{K}_n(g,\alpha)}{\beta(g)}
\times{T}\exp\Bigl[
\int_{\bar{g}(Q^2)}^{g}dg' \frac{\hat{\gamma}_n(g')}{\beta(g')}
\Bigr]~,\nonumber
\eea
with $\hat{\gamma}_n$ and $\vec{K}_n$ being the hadronic anomalous dimension
matrix and the off-diagonal element representing the mixing between
the photon and hadronic operators, respectively.  
The coefficient functions are expanded up to the one loop level as (see Eq.(3.15) of
Ref.\cite{SU1})
\bea
\vec{C}_n(1,\bar{g}(Q^2))&=&
\left(
\begin{array}{c}
C^\psi_n (1,\bar{g}(Q^2))\\
C^G_n (1,\bar{g}(Q^2))\\
C^{NS}_n (1,\bar{g}(Q^2))\\
\end{array}
\right)
=\left(
\begin{array}{c}
\langle e^2\rangle\left(1+\frac{\alpha_s(Q^2)}{4\pi}B_\psi^n\right)\\
\langle e^2\rangle\frac{\alpha_s(Q^2)}{4\pi}B_G^n\\
1+\frac{\alpha_s(Q^2)}{4\pi}B_{NS}^n\\
\end{array}
\right)~,\nn\\
C_n^\gamma(1,\bar{g},\alpha)&=&\frac{\alpha}{4\pi}
3n_f\langle e^4\rangle B_n^\gamma~.
\label{coeff-fn}
\eea

It is noted that, in Eq.(\ref{mom-ope}), the $P^2$-dependence only resides in 
$\vec{A}_n(Q_0^2;P^2)$, the photon matrix elements of the hadronic operators. 
When the photon state
becomes far off-shell and $P^2$ approaches $Q_0^2$, $\vec{A}_n(Q_0^2;P^2)$ are considered 
to be point-like  and can be evaluated perturbatively. 
Indeed, the point-like component has been calculated in the 
${\overline {\rm MS}}$
scheme~\cite{MvanN}, and  we  get in the leading order,
 \bea
\vec{A}_n(Q_0^2;P^2=Q_0^2)&\equiv&\vec{A}_n^{(0)}~\nn\\
&=&\Bigl(\langle e^2\rangle, 0, \langle e^4\rangle-\langle e^2\rangle^2   \Bigr)\nn\\
&&\times 12n_f\left[\frac{n-1}{n(n+1)}\sum_{k=1}^n \frac{1}{k}
+\frac{4}{(n+1)^2}-\frac{1}{n^2}  -\frac{1}{n}\right]~.\label{PointlikeA}
\eea
For an arbitrary $P^2$ in the range $0\leq P^2 \leq Q_0^2$, we divide
$\vec{A}_n(Q_0^2;P^2)$ into two pieces such that $\vec{A}_n(Q_0^2;P^2)=\vec{\widetilde
A}_n(Q_0^2;P^2)+\vec{A}_n^{(0)}$ with
\bea
\vec{\widetilde A}_n(Q_0^2;P^2)&\equiv&\Bigl(\vec{A}_n(Q_0^2;P^2)-
 \vec{A}_n^{(0)} \Bigr)\nn\\
&=&\Bigl({\widetilde A}_n^\psi(Q_0^2;P^2), {\widetilde A}_n^G(Q_0^2;P^2),{\widetilde
A}_n^{NS}(Q_0^2;P^2)\Bigr)~.
\eea 
Note that  $\vec{\widetilde A}_n(Q_0^2;P^2)$ contain nonperturbative 
contributions (i.e., hadronic components) when $P^2$ is in the range  $0\leq P^2 \leq Q_0^2$, and
satisfy the following boundary condition by definition:
\bea
\vec{\widetilde A}_n(Q_0^2;P^2=Q_0^2)=0~. \label{Q0-BC}
\eea

Then following the same procedures as we did
in Ref.\cite{SU1}, we obtain the formula for the $n$-th moment of $g_1^\gamma$
up to the NLO in QCD, 
\begin{eqnarray}
\int_0^1 dx x^{n-1}g_1^\gamma(x,Q^2,P^2)&=&
\frac{\alpha}{4\pi}\frac{1}{2\beta_0}\hspace{-0.0cm}
\biggl[
\sum_{i=+,-,NS}
L_i^n
\frac{4\pi}{\alpha_s(Q^2)}
\Bigl\{1-\left(\frac{\alpha_s(Q^2)}{\alpha_s(Q_0^2)}\right)^{d_i^n+1}
\Bigr\}\nonumber\\
&&\hspace{-3.5cm}+
\sum_{i=+,-,NS}{\cal A}_i^n\Bigl\{1-\left(\frac{\alpha_s(Q^2)}
{\alpha_s(Q_0^2)}\right)^{d_i^n}\Bigr\}
+\sum_{i=+,-,NS}{\cal B}_i^n\Bigl\{1-\left(\frac{\alpha_s(Q^2)}
{\alpha_s(Q_0^2)}\right)^{d_i^n+1}\Bigr\}\nonumber\\
&&\hspace{-2.5cm}+\ {\cal C}^n 
+2\beta_0\vec{\widetilde A}_n(Q_0^2;P^2)
\cdot \sum_{i=+,-,NS}P_i^n\vec{C}_n(1,0)\left(\frac{\alpha_s(Q^2)}
{\alpha_s(Q_0^2)}\right)^{d_i^n}\ \biggr]~,
\label{master}
\end{eqnarray}
which is applicable for any values of the target mass $P^2$ in the range $0\leq P^2 \leq
Q_0^2$. 
Here the coefficients $L_i^n$, ${\cal A}_i^n$ and ${\cal B}_i^n$ 
are computed from the one- and two-loop anomalous dimensions
together with one-loop coefficient functions and their explicit expressions 
are given in Ref.\cite{SU1}. The coefficient ${\cal C}^n$ is expressed as
\be
{\cal C}^n=2\beta_0\Bigl(3n_f\langle e^4\rangle B_\gamma^n+\vec{A}_n^{(0)}\cdot 
\vec{C}_n(1,0)\Bigr)~. \label{C^n}
\ee
The exponents $d_i^n$
are given by $d_i^n=\lambda_i^n/2\beta_0$ ($i=+,-,NS$) where $\lambda_i^n$ 
are the eigenvalues of the one-loop anomalous dimension matrix $\hat{\gamma}_n^{(0)}$,
which is expanded as $\hat{\gamma}_n^0 =\sum_{i}\lambda_i^n P_i^n$ with $P_i^n$
being the projection operators (see Appendix A of Ref.\cite{SU1} for more detail). 
It is noted that all the terms in the square brackets except the last one with 
$\vec{\widetilde A}_n(Q_0^2;P^2)$ are calculable by the perturbative QCD. 
Eq.(\ref{master}) is the master formula which can be
applied  to the  kinematical region where $Q_0^2$ satisfies the condition
(\ref{KineRegionQ0})  and $P^2$ takes any values between 0 and $Q_0^2$. 
When $P^2$ approaches $Q_0^2$, the last term with $\vec{\widetilde A}_n(Q_0^2;P^2)$ 
vanishes, and we recover our previous result (Eq.(3.16) of Ref.\cite{SU1}) for 
the $n$-th moment of $g_1^\gamma(x,Q^2,P^2)$ which is applicable 
to the case $\Lambda^2 \ll P^2 \ll Q^2$.

Now let us study  the first moment in the leading order. 
To this end we take the $n\rightarrow 1$ limit of the formula 
(\ref{master}). As shown in ref.\cite{SU1}, the terms proportional to 
$L_i^n$, ${\cal A}_i^n$ and ${\cal B}_i^n$ all vanish
in the $n=1$ limit. But the last two terms in (\ref{master}) 
contribute to the first moment. Since $B_\gamma^{n=1}=0$ and ${\vec{C}}_{n=1}(1,0)=
(\langle e^2\rangle, 0,1)^{\rm T}$ in Eq.(\ref{coeff-fn}), we find from
Eqs.(\ref{PointlikeA}) and  (\ref{C^n})
\be
{\cal C}^{n=1}=-24\beta_0 n_f\langle e^4\rangle~.
\ee
From the explicit expression of the one-loop anomalous dimension matrix 
$\hat{\gamma}_n^{(0)}$ at $n=1$ (see Appendix B of Ref.\cite{SU1}), we compute the
eigenvalues $\lambda_i^{n=1}$ ($i=+,-,NS$) and the corresponding projection operators
$P_i^{n=1}$,  and  find 
\be
\sum_{i=+,-,NS}P_i^{n=1}\vec{C}_{n=1}(1,0)\left(\frac{\alpha_s(Q^2)}
{\alpha_s(Q_0^2)}\right)^{d_i^{n=1}} =(\langle e^2\rangle, 0,1)^{\rm T}~.
\ee
Thus we obtain for the first moment of $g_1^\gamma(x,Q^2,P^2)$ up to the order of $\alpha$
\bea
\hspace{-0.5cm}
\int_0^1 dx g_1^\gamma(x,Q^2,P^2)
&=&-\frac{3\alpha}{\pi}n_f\langle e^4\rangle
 +\frac{\alpha}{4\pi}\left(
\langle e^2\rangle \widetilde{A}_{n=1}^\psi(Q_0^2;P^2)+
\widetilde{A}_{n=1}^{NS}(Q_0^2;P^2)\right)~,\nonumber\\
\label{master-4}
\eea
which holds for the range of the target photon mass squared:
$0\leq P^2 \leq Q_0^2$. It is emphasized that apart from the QED coupling constant $\alpha$, the 
leading order of the first moment is ${\cal O}(1)$ not of order
$1/\alpha_s(Q^2)$,  which is the case for the general moments  of $g_1^\gamma$ with
$n>1$. 

In the above equation (\ref{master-4}), the last two terms will vanish 
as we go to higher $P^2\gg\Lambda^2$, while at $P^2=0$ they 
cancel out the first term  which arises from the pure QED point-like interaction, 
as we see from Eq.(\ref{1momentReal}).
Now we investigate the $P^2$-dependence of these two terms. 
An argument which leads to Eq.(\ref{1momentReal}) goes as follows~\cite{BASS}. 
Only the quark operators contribute to the first moment of $g_1^\gamma$, since 
gauge-invariant twist-two gluon and photon operators with spin one are absent. 
This remark is supported also by the expression of Eq.(\ref{master-4}).
The relevant $n=1$ quark operators $R_{n=1,\sigma}^{\psi}$ and $R_{n=1,\sigma}^{NS}$ 
(we added the Lorentz index $\sigma$) are, 
in fact, the flavor singlet 
$J_{5 \sigma}^{\psi}=\overline{\psi}\gamma_\sigma\gamma_5{\bf 1}\psi$ 
and nonsinglet $J_{5 \sigma}^{NS}=\overline{\psi}\gamma_\sigma\gamma_5(Q_{\rm ch}^2-\langle e^2\rangle{\bf 1})\psi$ axial
currents, respectively, where $Q_{\rm ch}$ (${\bf 1}$) is the $n_f\times n_f$
quark-charge (unit) matrix.
Consider the photon matrix element of these axial vector
currents, which are expressed as
\bea
\langle \gamma(l)|J_{5 \sigma}^k|\gamma(p)\rangle
=R^k_{\sigma\alpha\beta}(p,l)\epsilon^{*\beta}(l)\epsilon^{\alpha}(p)~, \qquad k=\psi, NS~,
\eea
where $p$ and $l$ are external photon momenta, and
$\epsilon^{\alpha}(p)$ and $\epsilon^{\beta}(l)$ are photon
polarization vectors as shown in Fig.1.
\begin{figure}
\begin{center}
\includegraphics[scale=0.4]{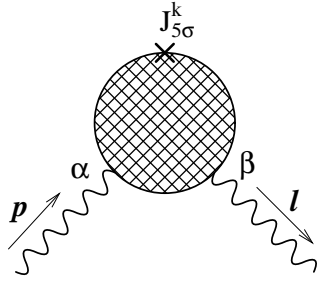}
\vspace{-0.5cm}
\caption{Vertex function $\displaystyle{R^k_{\sigma\alpha\beta}(p,l)}$ of the axial vector
current }
\end{center}
\end{figure}
Then the covariance, parity and crossing symmetry lead to
the following tensor decomposition for the vertex function \cite{Adler,BASS}:
\bea
\hspace{-0.5cm}
R^k_{\sigma\alpha\beta}(p,l)&=&
\epsilon_{\sigma\nu\alpha\beta}(p+l)^\nu\ A^k(s^2,p^2,l^2)
+\epsilon_{\sigma\alpha\nu\delta}p^\nu l^\delta\{B^k_1(s^2,p^2,l^2)p_\beta
\nonumber\\
&&\hspace{-0.3cm}
+B^k_2(s^2,p^2,l^2)l_\beta\}
+\epsilon_{\sigma\beta\nu\delta}p^\nu l^\delta
\{B^k_1(s^2,p^2,l^2)l_\alpha+B^k_2(s^2,p^2,l^2)p_\alpha\}
,\label{vertexfunction}
\eea
where $s^2=(p-l)^2$.
Imposing the current conservation, 
\bea
p^\alpha R^k_{\sigma\alpha\beta}=l^\beta R^k_{\sigma\alpha\beta}=0,
\eea
we obtain the following relations:
\bea
A^k(s^2,p^2,l^2)&=&p\cdot l \ B^k_1(s^2,p^2,l^2)+p^2 B^k_2(s^2,p^2,l^2)~,
\nonumber\\
&=&p\cdot l \ B^k_1(s^2,p^2,l^2)+l^2 B^k_2(s^2,p^2,l^2)~.
\eea
Now restricting ourselves to the case of the forward scattering where we put $p=l$ and
$s^2=(p-l)^2=0$,  we get
\bea
R^k_{\sigma\alpha\beta}(p,p)=
2\epsilon_{\sigma\nu\alpha\beta}
p^\nu A^k(0,p^2,p^2)=
\Bigl[2\epsilon_{\sigma\nu\alpha\beta}
p^\nu\Bigr] p^2 \left(B^k_1(0,p^2,p^2)+B^k_2(0,p^2,p^2)\right).\label{Bfunction1}
\eea
Since there is a relation
$R^k_{\sigma\alpha\beta}(p,p)=\Bigl[2\epsilon_{\sigma\nu\alpha\beta}p^\nu\Bigr]
\langle \gamma(p)|R_{n=1}^k(\mu^2)|\gamma(p)\rangle$, the matrix element
$\langle \gamma(p)|R_{n=1}^k(\mu^2)|\gamma(p)\rangle$ is proportional to $p^2(=-P^2)$.
Hence we conclude from Eq.(\ref{BasicFormula}) that 
the first moment of $g_1^\gamma$ is expressed in the form as
\bea
\int_0^1 dx g_1^\gamma(x,Q^2,P^2)
=P^2\widetilde{B}(P^2)\equiv B(P^2)~.\label{Bfunction}
\eea
At $P^2=0$ we have the vanishing sum rule (\ref{1momentReal}),
unless there appears a massless pole like a Nambu-Goldstone boson
in $\widetilde{B}(P^2)$. 

In order to examine the explicit $P^2$ dependence of the hadronic terms
$\widetilde{A}_{n=1}^\psi(Q_0^2;P^2)$ and $\widetilde{A}_{n=1}^{NS}(Q_0^2;P^2)$
in Eq. (\ref{master-4}), 
let us adopt the VMD model~\cite{JJS}.  
In this model the photon matrix element of the axial vector current $J_{5\sigma}$
is given by
\bea
\langle \gamma(p)\vert J_{5\sigma}(\mu)\vert\gamma(p)\rangle
=4\pi\alpha
\int_0^\infty dm^2\rho(m^2)\left(\frac{1}{m^2+P^2}\right)^2
\langle V(m)\vert J_{5\sigma}(\mu)\vert V(m)\rangle~,
\eea
where $\vert V(m)\rangle$ are the vector meson states with
mass $m$, and $\rho(m^2)$ is the spectral function for the generalized
vector dominance model \cite{SS}, which can be written as
\bea
\rho(m^2)=\sum_{V=\rho,\omega,\phi}\left(\frac{m_V^2}{f_V}
\right)^2\delta(m^2-m_V^2)+\widetilde{\rho}(m^2)~.
\eea
The first term corresponds to  the contribution from the 
low-lying vector mesons, $\rho$, $\omega$, $\phi$, and   
can be interpreted as the double-pole singularities
$\delta'(m^2-m_V^2)$ in the dispersion integral discussed in
Refs. \cite{BJ,IW,GASTS}. The second term $\widetilde{\rho}$ 
refers to the continuous part of the spectral function. Here we assume the dominance
of the low-lying vector mesons. 
Remembering that $\widetilde{A}_{n=1}^k(Q_0^2;P^2)$ with $k=\psi, NS$ represent 
hadronic components of the 
matrix elements  $\langle \gamma(p)|R_{n=1}^k(Q_0^2)|\gamma(p)\rangle$ 
and that the operators $R_{n=1}^k$ correspond to the axial vector currents, 
we expect that $\widetilde{A}_{n=1}^\psi(Q_0^2;P^2)$ and
$\widetilde{A}_{n=1}^{NS}(Q_0^2;P^2)$ may be expressed as
\bea
\widetilde{A}_{n=1}^\psi(Q_0^2;P^2)&=&
\sum_{V=\rho,\omega,\phi}
\frac{\langle e^2\rangle}{f_V^2}\left(\frac{m_V^2}{m_V^2+P^2}\right)^2
\langle V| J_{5}(Q_0^2)| V\rangle~,
\nonumber\\
\widetilde{A}_{n=1}^{NS}(Q_0^2;P^2)&=&
\sum_{V=\rho,\omega,\phi}
\frac{(\langle e^4\rangle-\langle e^2\rangle^2)}{f_V^2}
\left(\frac{m_V^2}{m_V^2+P^2}\right)^2
\langle V| J_{5}(Q_0^2)| V\rangle~,\label{vmd-tildeA}
\eea
where $\langle V| J_{5}(Q_0^2)| V\rangle$ is a common factor the magnitude of which 
will be determined later.
In Eq.(\ref{vmd-tildeA}), we have taken the viewpoint that the photon matrix
elements of the axial vector currents arise dominantly from the quark-loop
diagrams so that the flavor nonsinglet part
$\widetilde{A}_{n=1}^{NS}$ is proportional to
${\rm Tr}(Q_{\rm ch}^2(Q_{\rm ch}^2-\langle e^2\rangle{\bf 1}))=3n_f
\left(\langle e^4\rangle-\langle e^2\rangle^2\right)$, while 
the singlet part $\widetilde{A}_{n=1}^{\psi}$ is proportional 
to ${\rm Tr}(Q_{\rm ch}^2{\bf 1})=3n_f
\langle e^2\rangle$.

\begin{figure}
\begin{center}
\includegraphics[scale=0.4]{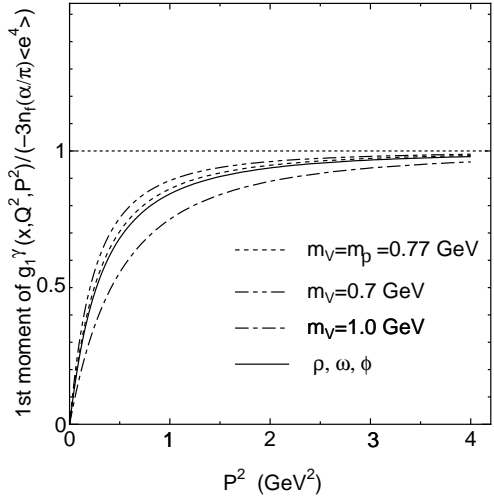} 
\caption{The $P^2$-dependence of the first moment  of $g_1^\gamma(x,Q^2,P^2)$
in units of $-3n_f(\alpha/\pi)\langle e^4\rangle$. 
The solid line shows the case of the three low-lying vector meson ($\rho$,
$\omega$,  $\phi$) dominance, while the short-dashed line is the result of the 
$\rho$ meson dominance. 
\label{PdepFirst}}
\end{center}
\end{figure}
In order for $\widetilde{A}_{n=1}^\psi(Q_0^2;P^2)$ and
$\widetilde{A}_{n=1}^{NS}(Q_0^2;P^2)$ to
satisfy the boundary condition (\ref{Q0-BC}), we either subtract the values
at $P^2=Q_0^2$ or introduce a sharp cutoff at $P^2=Q_0^2$. 
Here we  are considering   the case where
$Q_0^2$ is sufficiently large ($Q_0^2\gg\Lambda^2 $) and thus the subtraction terms are
negligible. On the other hand, the vanishing of the first moment of $g_1^\gamma$ at
$P^2=0$  gives a condition on $\langle V| J_{5}(Q_0^2)| V\rangle$. We find from 
Eqs.(\ref{master-4}) and (\ref{vmd-tildeA}), 
\bea
\sum_{V=\rho,\omega,\phi}C^{(V)}(Q_0^2)=1~,\qquad {\rm with} \quad
C^{(V)}(Q_0^2)\equiv\frac{1}{12n_f}\frac{\langle V|
J_{5}(Q_0^2)| V\rangle}{f_V^2}~.\label{CvCondi}
\eea
Since $Q_0^2$ is sufficiently large,  we expect that 
the $Q_0^2$ dependence of $C^{(V)}(Q_0^2)$ is 
rather mild, and hence we neglect the $Q_0^2$ dependence of 
$C^{(V)}(Q_0^2)$  from now on and write simply as $C^{(V)}$.
Then the sum rule is now written as
\bea
\int_0^1 dx g_1^\gamma(x,Q^2,P^2)
=-\frac{3\alpha}{\pi}n_f\langle e^4\rangle
 +\frac{3\alpha}{\pi}n_f\langle e^4\rangle \sum_{V=\rho,\omega,\phi}
C^{(V)}\left(\frac{m_V^2}{m_V^2+P^2}\right)^2~.
\label{VMD-sum-rule-LO}
\eea
In other words, the vertex function $B(P^2)$ in Eq.(\ref{Bfunction}) is expressed as
\bea
B(P^2)=P^2\widetilde{B}(P^2)=-\frac{3\alpha}{\pi}
n_f\langle e^4\rangle P^2\sum_VC^{(V)}
\frac{(2m_V^2+P^2)}{(m_V^2+P^2)^2}
~,\label{${B}(P^2)$}
\eea
and we see that
$\widetilde{B}(P^2)$ is actually non-singular at $P^2=0$.

In Fig. 2 we have plotted the $P^2$-dependence of
the first moment  of $g_1^\gamma(x,Q^2,P^2)$ given in Eq.(\ref{VMD-sum-rule-LO}).
The solid line shows the case when  
the contributions of the three
low-lying vector mesons, $\rho$, $\omega$, $\phi$, are taken into account,  
with  their masses 
$m_\rho=0.77$ GeV, $m_\omega=0.78$ GeV, $m_\phi=1.02$ GeV 
\cite{PDG}. We adopt the values 
$f_\rho^2/4\pi\simeq 2.36$, $f_\omega^2/4\pi\simeq 20$, 
$f_\phi^2/4\pi\simeq 12$ \cite{FGS} which lead to $C^{(\rho)}=0.76$, 
$C^{(\omega)}=0.09$, and $C^{(\phi)}=0.15$. 
Also shown are the cases of the simple pole saturation by $\rho$ meson:
$C_{n=1}^\rho=1$ and  (i) $m_\rho=0.77$ GeV (short-dashed line),  and some variation 
of $\rho$ mass, (ii) 
$m_V=0.7$ GeV (2dot-dashed) and (iii) $m_V=1.0$ GeV (dot-dashed). 
We see that the $\rho$-dominance is a good
approximation to the case where all the three low-lying vector mesons $\rho$, $\omega$, $\phi$
are included.

Next we pursue the QCD corrections of order $\alpha_s$ to the first 
moment of $g_1^\gamma(x,Q^2,P^2)$ for an arbitray target mass $P^2$ in the 
range $0\leq P^2 \leq Q_0^2$. We now take into account
the $\alpha_s$ corrections to the coefficient functions in 
Eq.(\ref{coeff-fn}) as well as
the ${\cal O}(\alpha_s)$ terms in the evolution factor 
${M}_n(Q^2/Q_0^2,\bar{g}(Q^2))$ 
in Eq.(\ref{FactorMn}) (see Ref.\cite{SU1} for detail). We can use the 
same $\widetilde{A}_{n=1}^{\psi}(Q_0^2;P^2)$ and 
$\widetilde{A}_{n=1}^{NS}(Q_0^2;P^2)$
given in Eq.(\ref{vmd-tildeA}) and we get
\bea
&&\int_0^1 dx g_1^\gamma(x,Q^2,P^2)\nonumber\\
&&=-\frac{3\alpha}{\pi}n_f \langle e^4 \rangle \left(1-  \frac{\alpha_s(Q^2)}{\pi} \right)
+\frac{3\alpha}{\pi}n_f \langle e^2 \rangle^2
\frac{2n_f}{\beta_0}\left(\frac{\alpha_s(Q_0^2)}{\pi}
-\frac{\alpha_s(Q^2)}{\pi}\right)\nonumber\\
&&\quad +\frac{\alpha}{4\pi}\langle e^2\rangle
\tilde{A}_{n=1}^{\psi}(Q_0^2;P^2)
\left\{1-\frac{\alpha_s(Q^2)}{\pi}
-\frac{2n_f}{\beta_0}\left(\frac{\alpha_s(Q_0^2)}{\pi}
-\frac{\alpha_s(Q^2)}{\pi}\right) 
\right\}\nonumber\\
&&\quad+\frac{\alpha}{4\pi}
\tilde{A}_{n=1}^{NS}(Q_0^2;P^2)\left(
1-\frac{\alpha_s(Q^2)}{\pi}\right)
. \quad \label{first-alpha}
\eea
At $P^2=Q_0^2$, we have $\tilde{A}_{n=1}^{\psi}(Q_0^2=P^2;P^2)=
\tilde{A}_{n=1}^{NS}(Q_0^2=P^2;P^2)=0$, and we recover the previous result
(\ref{1momentVirtual}). For the real photon target ($P^2=0$), 
we  obtain from Eq.(\ref{vmd-tildeA}) and the condition (\ref{CvCondi}),
\bea
\tilde{A}_{n=1}^{\psi}(Q_0^2;P^2=0)=12n_f\langle e^2\rangle,\quad
\tilde{A}_{n=1}^{NS}(Q_0^2;P^2=0)=12n_f
\left(\langle e^4\rangle-\langle e^2\rangle^2\right)~,\label{real-A1}
\eea
and thus we see
that the first moment of $g_1^\gamma$ to the order $\alpha_s$ indeed 
vanishes for the real photon target.
If we use the vertex function ${B}(P^2)$ given in Eq.(\ref{${B}(P^2)$}), 
the first moment (\ref{first-alpha}) can be rewritten as
\bea
&&\int_0^1 dx g_1^\gamma(x,Q^2,P^2)\nonumber\\
&&=B(P^2)\left[
\left(1-\frac{\alpha_s(Q^2)}{\pi}\right)
-\frac{n_f\langle e^2\rangle^2}{\langle e^4\rangle}
\frac{2}{\beta_0}
\left(\frac{\alpha_s(Q_0^2)}{\pi}-\frac{\alpha_s(Q^2)}{\pi}\right)\right]~.
\eea
Note that the order $\alpha_s$ QCD effect is factorizable and amounts to 
reduce the leading order result by about 7$\%$ for $Q_0^2=1.0$ GeV$^2$ and 
$Q^2=30$ GeV$^2$.
This analysis would be easily extended to the case for the order $\alpha_s^2$ 
QCD effect on the first moment \cite{SUU}.
We also note here that there has been an analysis of the first moment
sum rule of $g_1^\gamma(x,Q^2,P^2)$ for the intermediate values of $P^2$
based on the more genereral principle like chiral symmetry and making a
connection with the off-shell radiative couplings of the pseudo-scalar 
mesons \cite{Shore}.

Finally we investigate the $x$-dependence of the structure function
$g_1^\gamma(x,Q^2,P^2)$ for an arbitrary value of $P^2$ between 0 and $Q_0^2$. 
Once we know all the moments, we can perform the inverse Mellin transform of 
the moments to get  $g_1^\gamma$ as a function of $x$.
The $n$-th moment of $g_1^\gamma(x,Q^2,P^2)$ up to the NLO in QCD is given by 
the formula (\ref{master}). All the quantities in there  are already known
and are collected in Ref.\cite{SU1},  except for
${\widetilde A}_n^\psi(Q_0^2;P^2)$, ${\widetilde A}_n^G(Q_0^2;P^2)$ and ${\widetilde
A}_n^{NS}(Q_0^2;P^2)$. Since photon does not 
couple to gluon field directly,  the contribution of the term with
${\widetilde A}_n^G(Q_0^2;P^2)$ 
is expected to be much smaller than those with ${\widetilde A}_n^\psi(Q_0^2;P^2)$ and ${\widetilde
A}_n^{NS}(Q_0^2;P^2)$ and thus  we assume ${\widetilde A}_n^G(Q_0^2;P^2)=0$.
To estimate the $P^2$-dependence of  ${\widetilde
A}_n^\psi(Q_0^2;P^2)$ and
${\widetilde A}_n^{NS}(Q_0^2;P^2)$, we adopt the VMD model again. 
We have seen in Fig.\ref{PdepFirst} that as regards the $P^2$-dependence of the first moment of $g_1^\gamma$, the $\rho$-dominance is a good
approximation to the case where all the three low-lying vector mesons $\rho$, $\omega$, $\phi$
are included. Therefore, we assume the $\rho$-dominance and consider only
the contribution  of $\rho$ meson taking $m_V=m_\rho$.
Now recalling that ${\widetilde A}_n^\psi(Q_0^2;P^2)$ and ${\widetilde
A}_n^{NS}(Q_0^2;P^2)$ satisfy the boundary condition (\ref{Q0-BC}) at $P^2=Q_0^2$, we  take the  following forms for $0\leq P^2 \leq Q_0^2$:
\bea
\widetilde{A}_n^\psi(Q_0^2;P^2)&=&12n_f\langle e^2\rangle
\left[\left(\frac{m_\rho^2}{m_\rho^2+P^2}\right)^2
-\left(\frac{m_\rho^2}{m_\rho^2+Q_0^2}\right)^2\right]\nonumber\\
&&\hspace{3cm}\times\left(1-\left(\frac{m_\rho^2}{m_\rho^2+Q_0^2}\right)^2\right)^{-1}
\times \tilde{f}(n)~,\\
\widetilde{A}_n^{NS}(Q_0^2;P^2)
&=&12n_f\left(\langle e^4\rangle -\langle e^2\rangle^2\right)
\left[\left(\frac{m_\rho^2}{m_\rho^2+P^2}\right)^2
-\left(\frac{m_\rho^2}{m_\rho^2+Q_0^2}\right)^2\right]\nonumber\\
&&\hspace{3cm}\times\left(1-\left(\frac{m_\rho^2}{m_\rho^2+Q_0^2}\right)^2\right)^{-1}
\times \tilde{f}(n)~, 
\eea
where
we have inserted a multiplication factor $(1-(m_\rho^2/(m_\rho^2+Q_0^2))^2)^{-1}$ 
so that $\widetilde{A}^{\psi}$ and $\widetilde{A}^{NS}$ with $n=1$ fulfill the
conditions (\ref{real-A1}) at $P^2=0$.
Here $\tilde{f}(n)$ is the Mellin transform of the quark parton
distribution function $f(x)$ inside the vector meson,  which we assume 
to be a binomial function as follows:
\bea
f(x)=B(p,q)^{-1} x^{p-1}(1-x)^{q-1}, \quad \tilde{f}(n)=
\int_0^1 x^{n-1} f(x)dx,\quad \tilde{f}(n=1)=1~.
\eea
Although there might be more sophisticated VMD inputs \cite{GASTS,AFG},
we consider the simplest case with $p=q=2$.

\begin{figure}
\begin{center}
\includegraphics[scale=0.4]{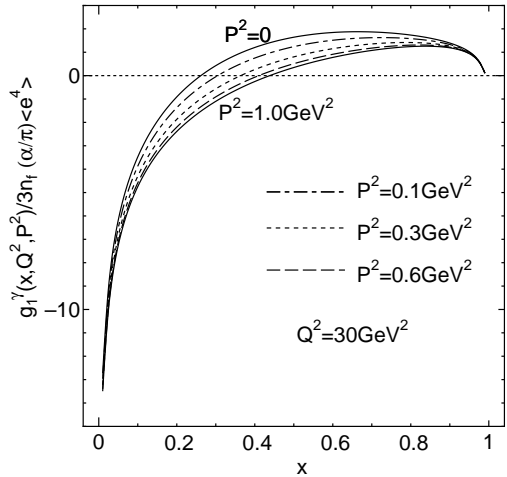}
\caption{The $P^2$-dependence of $g_1^\gamma(x,Q^2,P^2)$ for the case of $Q^2=$ 30 GeV$^2$, $Q_0^2=$
1.0 GeV$^2$, 
$n_f=3$ and $\Lambda=0.2$ GeV. }
\end{center}
\end{figure}

In Fig. 3, we have plotted  $g_1^\gamma(x,Q^2,P^2)$ 
in units of $3n_f(\alpha/\pi)\langle e^4\rangle$ for 
various values of $P^2$. We took
$P^2=$ 0, 0.1, 0.3, 0.6, 1.0 GeV$^2$ with $Q^2=$ 30 GeV$^2$, $Q_0^2=$ 1.0 GeV$^2$, 
$n_f=3$ and $\Lambda=0.2$ GeV. 
Note that we can read off the tendency
where the first moment sum rule vanishes for the  real photon ($P^2=0$), and
then turns to be more negative 
as the mass squared of the virtual photon $P^2$ increases.

Here we also note that the $g_1^\gamma$ of real and virtual photons have been
studied in NLO QCD using the positivity constraints in \cite{GRS}, where
the authors presented the maximal and minimal values of $g_1^\gamma$ for
real ($P^2=0$) and virtual ($P^2=1$GeV$^2$) photons, which appear to be 
consistent with our present analysis.

In order to make sure that our analysis is stable under the change of the
renormalization scale $Q_0^2$, we show in Fig. 4 the $Q_0^2$-dependence
of $g_1^\gamma(x,Q^2,P^2)$ in units of $3n_f(\alpha/\pi)\langle e^4\rangle$
for the case of a real photon ($P^2=0$),  with
$Q^2=$30GeV$^2$, $n_f=3$ and $\Lambda=0.2$ GeV. Three curves with
$Q_0^2=$ 0.75, 1.00, 1.25 GeV$^2$ almost overlap in the whole $x$ 
region and we see that there
appears no sizable dependence on $Q_0^2$. 

In summary, we have investigated in QCD the transition of the polarized photon
structure function $g_1^\gamma$ when the target photon shifts from on-shell to
far off-shell region  up to the NLO. The  first moment of $g_1^\gamma$ which 
vanishes for the real photon, turns to a negative value when the target photon
becomes off-shell. Although our estimate of the non-perturbative effects of
the photon matrix elements relies on the VMD model, we have studied 
the explicit $P^2$-dependence not only for the first moment sum rule but also 
for the structure function $g_1^\gamma(x,Q^2,P^2)$, in particular, as a 
function of $x$.
It turns out that the results are not so sensitive to the choice of the 
renormalization scale.

\vspace{0.5cm}
This work is supported in part by Grant-in-Aid for Scientific Research 
from the Ministry of Education, Culture, Sports, Science and Technology,
Japan No.18540267.

\begin{figure}
\begin{center}
\includegraphics[scale=0.4]{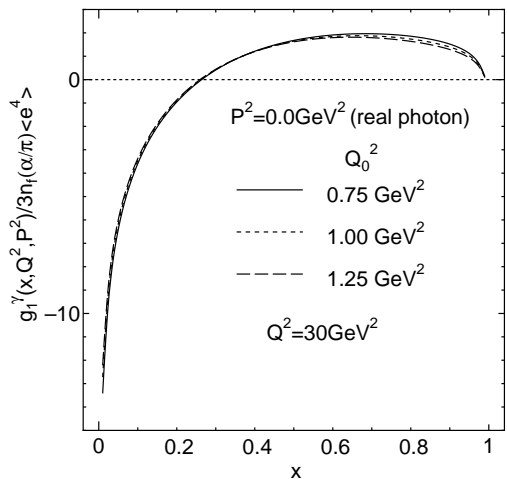}
\caption{The $Q_0^2$-dependence of $g_1^\gamma(x,Q^2,P^2)$ for the real photon
with
$Q^2=$30GeV$^2$, $n_f=3$ and $\Lambda=0.2$ GeV. }
\end{center}
\end{figure}

\end{document}